\documentclass[aip,jcp,reprint,floatfix]{revtex4-1}

\usepackage{graphicx}
\usepackage{xcolor}
\usepackage{amsmath}
\usepackage[colorlinks=true, urlcolor=blue, citecolor=blue, linkcolor=blue]{hyperref}

\begin{document}

\title{\sf{This is an author-created postprint. The version of record appeared in\\ Low Temp. Phys. \textbf{46}, 331--337 (2020); Fiz. Nizk. Temp. \textbf{46}, 402--409 (2020); \href{https://doi.org/10.1063/10.0000863}{https://doi.org/10.1063/10.0000863}. \textcopyright 2020 Authors} \hrule \vskip 20pt \huge{Angular magnetic-field dependence of vortex matching in pinning lattices fabricated by focused or masked helium ion beam irradiation of superconducting YBa$_2$Cu$_3$O$_{7-\delta}$ thin films}}

\author{B.~Aichner}

\author{K.~L.~Mletschnig}

\affiliation{University of Vienna, Faculty of Physics, Electronic Properties of Materials, Boltzmanngasse 5, A-1090, Wien, Austria}

\author{B. M\"uller}
\author{M. Karrer}
\affiliation{Physikalisches Institut and Center for Quantum Science (CQ) in LISA$^+$, Universit\"at T\"ubingen, Auf der Morgenstelle 14, D-72076 T\"ubingen, Germany}

\author{M.~Dosmailov}
\affiliation{Johannes-Kepler-University Linz, Institute of Applied Physics, Altenbergerstrasse 69, A-4040 Linz, Austria}
\affiliation{Al-Farabi Kazakh National University, Almaty, Kazakhstan}

\author{J.~D.~Pedarnig}

\affiliation{Johannes-Kepler-University Linz, Institute of Applied Physics, Altenbergerstrasse 69, A-4040 Linz, Austria}

\author{R. Kleiner}

\author{D. Koelle}
\affiliation{Physikalisches Institut and Center for Quantum Science (CQ) in LISA$^+$, Universit\"at T\"ubingen, Auf der Morgenstelle 14, D-72076 T\"ubingen, Germany}

\author{W.~Lang}
\email[Corresponding author: ]{wolfgang.lang@univie.ac.at}

\affiliation{University of Vienna, Faculty of Physics, Electronic Properties of Materials, Boltzmanngasse 5, A-1090, Wien, Austria}

\begin{abstract}
The angular dependence of magnetic-field commensurability effects in thin films of the cuprate high-critical-temperature superconductor YBa$_{2}$Cu$_{3}$O$_{7-\delta}$ (YBCO) with an artificial pinning landscape is investigated. Columns of point defects are fabricated by two different methods of ion irradiation --- scanning the focused 30~keV ion beam in a helium ion microscope or employing the wide-field 75~keV He$^+$ beam of an ion implanter through a stencil mask. Simulations of the ion-target interactions and the resulting collision cascades reveal that with both methods square arrays of defect columns with sub-$\mu$m spacings can be created. They consist of dense point-defect clusters, which act as pinning centers for Abrikosov vortices. This is verified by the measurement of commensurable peaks of the critical current and related minima of the flux-flow resistance vs magnetic field at the matching fields. In oblique magnetic fields the matching features are exclusively governed by the component of the magnetic field  parallel to the axes of the columnar defects, which confirms that the magnetic flux is penetrated along the defect columns. We demonstrate that the latter dominate the pinning landscape despite of the strong intrinsic pinning in thin YBCO films.

\end{abstract}



\maketitle

\section{Introduction}

Most of the superconducting materials belong to the type II class, into which a magnetic field can penetrate as flux quanta $\Phi_0 = h/(2e)$, where $h$ is Planck's constant and $e$ the elementary charge. These flux quanta are known as Abrikosov vortices, whirls of the supercurrent that confine the magnetic flux into the cylindrical vortex core, a region with vanishing density of superconducting charge carrier pairs. In clean and isotropic bulk superconductors these vortices arrange themselves in a two-dimensional hexagonal lattice with the axes of the vortex cores oriented parallel to the external magnetic field.

In the cuprate high-temperature superconductors (HTSCs) the situation is more complex and vortices can exist in a large range of magnetic fields between a tiny lower critical field $B_{c1}(85\,\text{K}) \sim 2$\,mT and a high upper critical field $B_{c2}(85\,\text{K}) \sim 20$\,T at temperatures relevant for the present study and for magnetic fields orthogonal to the CuO$_2$ atomic layers. \cite{KOBA88} The high anisotropy of the HTSCs favors a decomposition of the cylindrical vortices into a stack of coupled ``pancake'' vortices, \cite{CLEM91} which can be visualized, e.g., in Bi$_2$Sr$_2$CaCu$_2$O$_8$. \cite{CORR19}

The arrangement of vortices in type-II superconductors can be tailored by the introduction of artificial defects as pinning sites for vortices. Those defects can be classified by their dimensionality and can severely disturb the native hexagonal vortex arrangement. Zero-dimensional (0D) point defects can be introduced as tiny non-superconducting impurities \emph{in-situ} during fabrication \cite{HAUG04} or by postprocessing with electron \cite{VOVK19} or light-ion irradiation of HTSCs. \cite{LANG10R} One-dimensional (1D) defects are commonly created by irradiation with swift heavy ions that produce amorphous channels with diameters of several nm, i.e., a few times the in-plane coherence length. They have been extensively investigated as a tool to enhance the critical current density. \cite{CIVA97R} Finally, grain boundaries and, in the prototypical HTSC YBa$_{2}$Cu$_{3}$O$_{7-\delta}$ (YBCO) also twin planes, can form two-dimensional (2D) defects that can pin vortices.

The dimensionality of the artificial defects is revealed by different angle-dependent behavior in tilted magnetic fields of superconducting properties like the critical current $I_c$, the vortex-flow resistance $R$, and $B_{c2}$. While point defects lead to a marginal angular dependence, randomly distributed yet parallel oriented 1D columnar defects cause narrow features in $I_c$, $R$, and magnetization vs field direction, centered around the magnetic field direction parallel to their symmetry axes. \cite{HOLZ93,PROZ94,SILH02} Similar observations hold for 2D defect planes in YBCO when the magnetic field is rotated through a direction that is oriented parallel to these planes. \cite{ROAS90}

In this work, we investigate artificial pinning lattices that are different in two aspects. First, they are neither strictly 0D or 1D, since they consist of dense point defects that form columnar defect clusters (CDs) with diameters at least one order of magnitude larger than the in-plane coherence length $\xi_{ab}(0) = 1.2$\,nm in YBCO. \cite{SEKI95} Second, these CDs are arranged in a periodic pattern that gives rise to commensurability effects at matching vortex and defect densities. Such commensurability effects have been primarily studied in metallic superconductors using arrays of holes (antidots) \cite{FIOR78,LYKO93,METL94,METL98,KEMM06,MISK10,MOSH11M} or magnetic dots \cite{VELE08R} in the material, nanogrooves, \cite{SHKL06,SHKL08,DOBR12} and superlattices \cite{DOBR18} but are also found in YBCO perforated with holes. \cite{CAST97}

The fabrication techniques of pinning arrays in our samples are based on the observation that irradiation of YBCO thin films with He$^+$ ions of moderate energy introduces point defects by displacing mainly oxygen atoms. This leads to a reduction of the transition temperature $T_c$, \cite{WANG95b} which can be well controlled by the ion fluence. \cite{SEFR01,LANG04,LANG10R,CYBA14a} By ion irradiation through a shadow mask \cite{LANG06a,LANG09,PEDA10,SWIE12,TRAS13,HAAG14,ZECH17a,ZECH18,ZECH18a} or using the focused ion beam of a He ion microscope (HIM) \cite{AICH19} an array of CDs can be created that acts as a pinning landscape for vortices.

Only few investigations have addressed the angular dependence of vortex commensurability effects in metallic superconductors with antidots \cite{METL98,WOMA19} and in YBCO thin films patterned with periodic CDs by ion irradiation. \cite{TRAS13,ZECH18} The purpose of this study is to explore whether the pinning landscapes created in YBCO by focused He$^+$ ion irradiation in a HIM act as 1D line-like pinning centers despite of consisting of 0D point defect clusters with inhomogeneous density.

\section{Experimental methods}

Epitaxial thin films of YBa$_2$Cu$_3$O$_{7-\delta}$ are grown on (100) MgO single-crystal substrates by pulsed-laser deposition using 248~nm KrF-excimer-laser radiation at a fluence of 3.2~J/cm$^2$. The thicknesses of the films used in this work are $t_z = (80 \pm 5)$~nm (sample SQ200) and $t_z = (210 \pm 10)$~nm (sample SQ500). The critical temperatures of the as-prepared films are $T_c \sim 90$~K with transition widths $\Delta T_c \sim 1$~K. The films are patterned by photolithography and wet chemical etching to form bridges with a length of $240~\mu \mathrm{m}$\ and a width of $w = 60~\mu \mathrm{m}$. Electrical contacts in a four-probe geometry are established on side arms of the bridges using sputtered Au pads with a voltage probe distance of $100\ \mu \mathrm{m}$.

\begin{figure*}[t]
\centering

\includegraphics*[width=\textwidth]{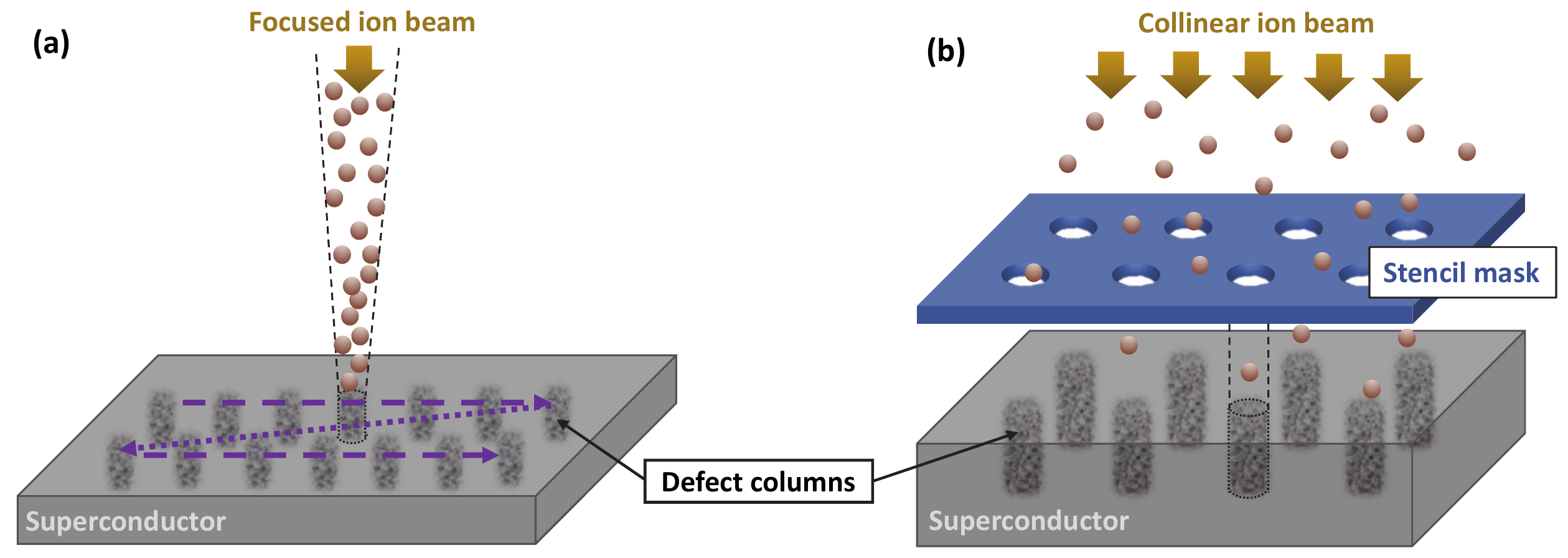}
\caption[]{Two different methods for patterning a YBCO film by He$^+$ ion irradiation: (a) Irradiation with a slightly defocused beam of a helium-ion microscope produces tailored columnar defect patterns by scanning the beam over the sample surface. The dark regions indicate the defect-rich, non-superconducting nano-cylinders. (b) Ion beam direct patterning by irradiating through a stencil mask creates a large number of columnar defects in a single step.}
\label{fig:irradiation}
\end{figure*}

In both samples, a tailored vortex pinning landscape was created by different methods of He$^+$ ion irradiation. Sample SQ200 was irradiated with an intentionally defocused ion beam in a HIM. The setup starts with adjusting the HIM settings to highest resolution and then changing the working distance (beam focus plane) so that the beam hits the sample surface with a nearly Gaussian fluence profile \cite{EMMR16} with a full width at half maximum (FWHM) of about 50~nm. Since the aperture angle of the ion beam is very small the ion beam hits the sample surface almost orthogonally. The method is described in detail elsewhere. \cite{AICH19}

By sequentially scanning the ion beam over the sample surface, a square lattice of columnar defects with $d = 200$\,nm spacings is created in the thin YBCO film in an overall area of approximately $200\,\mu{\rm m}\times 100\,\mu{\rm m}$. Every point is irradiated with 30 keV He$^+$ ions with a dwell time of 2.7\,ms and a beam current of 3\,pA, corresponding to $\sim\,51000$ He$^+$ ions/point. The method is sketched in Fig.~\ref{fig:irradiation}(a).

Sample SQ500 is patterned by masked ion beam structuring (MIBS) \cite{LANG06a} as sketched in Fig.~\ref{fig:irradiation}(b). A $2\,\mu$m-thick Si stencil mask is placed on top of the YBCO film and adjusted in an optical microscope with the help of marker holes. The mask is separated from the surface of the YBCO film by a circumferential spacer layer made of $1.5\,\mu$m-thick photoresist. The stencil mask is perforated with holes with diameters $D = (180 \pm 5)$~nm, arranged in a square array of $d = (500 \pm 2)\,$nm pitch, which covers the entire bridge. The stencil pattern is shadow projected onto the YBCO surface by irradiating the arrangement with a collinear 75\,keV He$^+$ ion beam, oriented orthogonal to the sample surface, in a commercial ion implanter (High Voltage Engineering Europa B. V.).

Electrical transport measurements are performed in a closed-cycle cryocooler with temperature control by a Cernox resistor, which has a negligible temperature reading error in moderate magnetic fields. \cite{HEIN98} The applied magnetic field $\mathbf{B_a}$ is supplied by a revolvable electromagnet with $\pm 1^\circ$ angular resolution and $B_a = |\mathbf{B_a}|$ is measured by a calibrated Hall probe mounted between the magnet's pole pieces. The Hall probe is connected to a LakeShore 475 gaussmeter, allowing for measurements of $B_a$ with a zero offset $< 10 \mu$T, and a reading accuracy $<0.1 \%$. The tilt angle $\alpha$ is defined as the angle between the surface normal of the YBCO film and the direction of $B_a$. The angle-dependent magnetoresistance measurements are performed in constant Lorentz force geometry, i. e., the magnetic field is always perpendicular to the current direction. For all measurements the current $I$ through the sample is generated by a constant-current source in both polarities to eliminate thermoelectric signals and the voltage $V$ is measured by a Keithley 2182A nanovoltmeter. The critical current $I_c(B_a)$ is determined from isothermal current-voltage ($I-V$) measurements with a voltage criterion of 100~nV, corresponding to $10\,\mu$V/cm. Since the $I-V$ characteristics of a superconductor are nonlinear the resistance curves presented below are defined as $R(B_a) = V(B_a)/I$ at a fixed $I$. Note that the absolute value of $R(B_a)$ is not important for our analyses.

\section{Results and discussion}

To compare the shapes of the artificial CD lattices, prepared by the two different irradiation methods, simulations of the defect distributions with the program package SRIM/TRIM \cite{ZIEG85M,ZIEG10} are performed. It computes the impact of ions on solids using a binary collision approximation of ion-atom and atom-atom collisions, and delivers the full collision cascades. However, ion channeling, thermal effects, diffusion, and recrystallization are not considered.

Details of the crystallographic structure are not considered in  SRIM/TRIM as it uses a Monte-Carlo method and assumes amorphous targets. For the spatial modulation of superconductivity, the Ginzburg-Landau coherence length is the relevant length parameter and therefore we have determined the average defect density within calculation cells of $2 \times 2 \times 2$~nm$^3$ --- a length scale of the order of the in-plane coherence length of YBCO. Note that the investigated point defect densities are below the amorphization limit and a comparison to an experimental visualization is hardly possible. Only by using a larger ion fluence, amorphous channels can be created and detected in cross-section scanning transmission electron microscopy images.\cite{MULL19}

The pinning potential for vortices is provided by a local suppression of $T_c$, which can be calculated from the defect density on the grounds of the pair-breaking theory of Abrikosov and Gor'kov. \cite{ABRI60} Since annealing effects are not considered in SRIM/TRIM and various other effects may lead to substantial uncertainty, a ``calibration'' curve relating the experimentally observed $T_c$ to the defect density from the simulations is established, using previous experimental $T_c$ values from full-area irradiation of thin YBCO films. \cite{LANG04} Details of this procedure are described elsewhere. \cite{MLET19}

\begin{figure*}[t]
\centering
\includegraphics*[width=\textwidth]{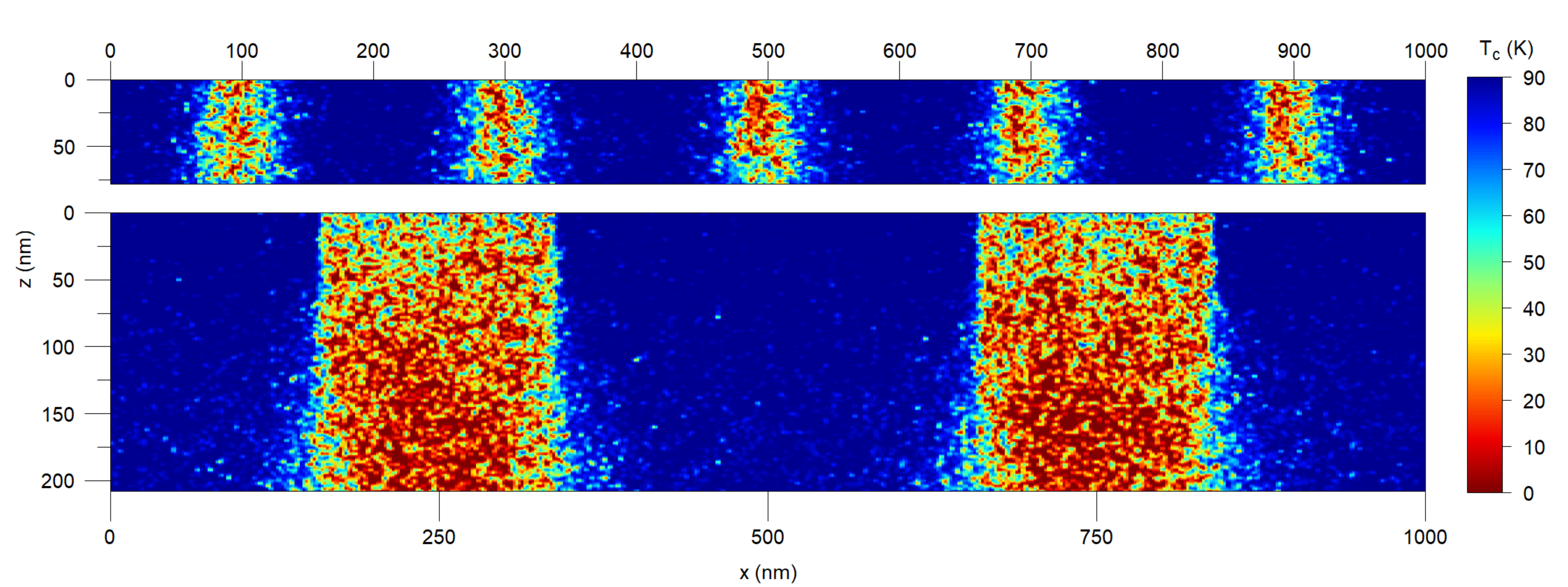}
\caption[]{Cross-sectional view of calculated local $T_c$ profiles within and around the defect columns produced by 50900 ions per dot with 30~keV energy and a Gaussian normal distribution with 50\,nm FWHM (sample SQ200, top panel) and 75~keV He$^+$ ion irradiation of YBCO with a fluence of $3\times10^{15}\textrm{~cm}^{-2}$ (sample SQ500, bottom panel). Both panels are displayed at the same scale for a comparison between the two samples.}
\label{fig:simu}
\end{figure*}

The resulting simulated cross-sectional $T_c$ profiles for the two samples SQ200 and SQ500 are presented in Fig.~\ref{fig:simu} at the same scale for comparison. Note that sample SQ200 (top panel) was irradiated with a slightly defocused He$^+$ ion beam with approximately Gaussian normal distributed fluence of FWHM~=~50\,nm, whereas the fluence was homogeneous in the irradiated parts of sample SQ500. Another important difference is the ion energy of 30\,keV for sample SQ200 and 75\,keV for sample SQ500.

In thin films with $t_z \le 80\,$nm, 30\,keV He$^+$ ion irradiation creates columns, within which $T_c$ is suppressed, that are clearly separated from each other at 200\,nm lattice spacing (Fig.~\ref{fig:simu}, top panel). The suppression of $T_c$ at the fringes of the CDs decays more gradually than for sample SQ500, which was irradiated by MIBS (Fig.~\ref{fig:simu}, bottom panel). Still, the cylindrical envelope of clusters with suppressed $T_c$ provides an efficient pinning landscape as will be discussed below.

Due to the larger penetration depth of the 75\,keV He$^+$ ions, CDs can be patterned into thicker YBCO films with MIBS, as demonstrated in Fig.~\ref{fig:simu}, bottom panel. However, the achievable lateral resolution for CD diameters degrades with increasing thickness of the film, as it can be noticed by the increasing diameter of the CD for film thicknesses larger than ~120\,nm. We note that increasing the ion energy would improve the resolution on the cost of a lower ion scattering cross section, which would demand a higher ion fluence.

Although a few dispersed defects are created also outside the CDs by lateral straggling of the incident ions and the secondary collision cascades, their impact on the zero-field electrical transport properties is marginal as demonstrated by the experimentally determined small reduction of the critical temperature $\Delta T_c = 2.6$\,K ($\Delta T_c = 4$\,K) in sample SQ200 (SQ500) after irradiation.

\begin{figure}[t]
\centering
\includegraphics[width=\columnwidth]{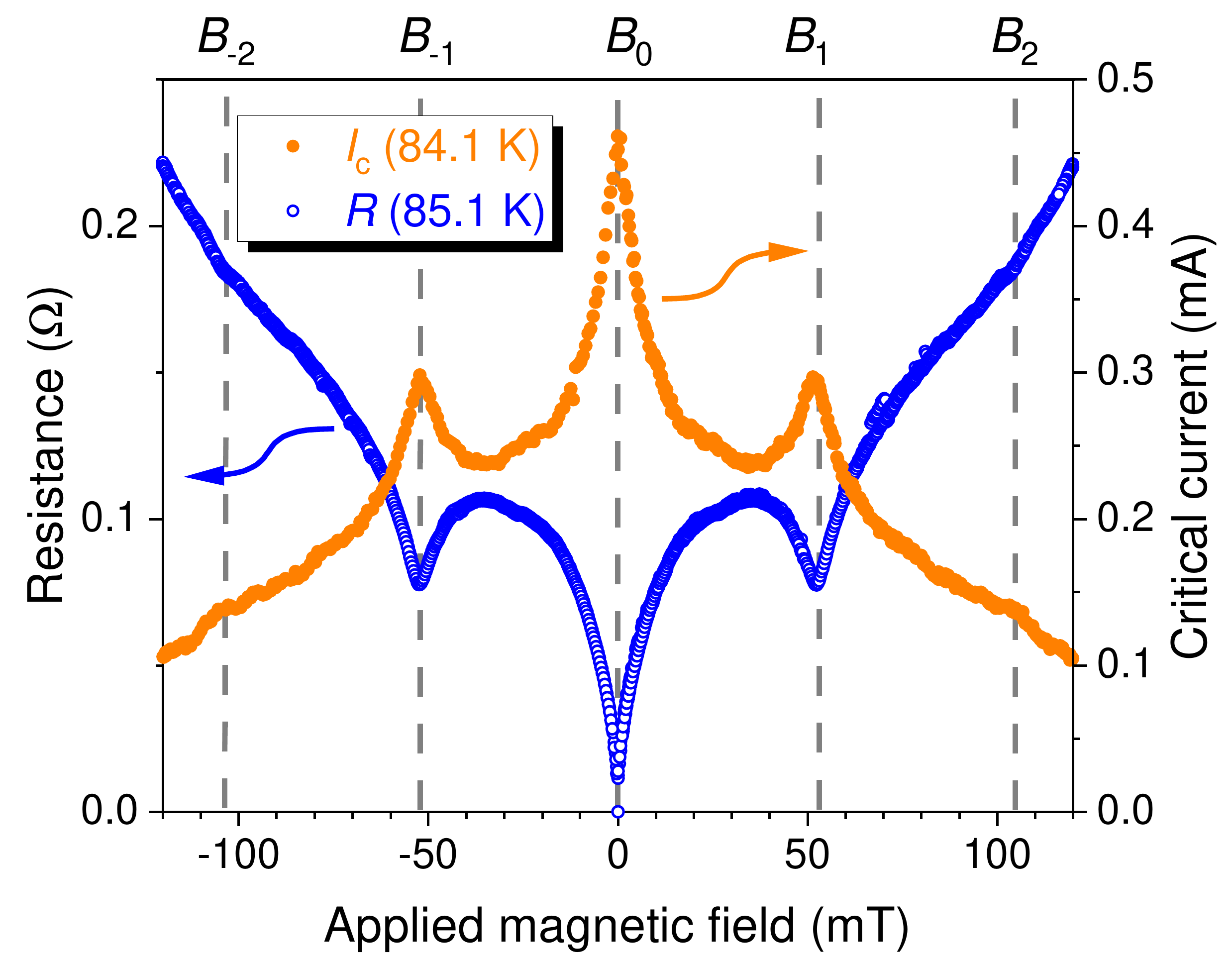}
\caption{Resistance $R$ ($I = 50\,\mu$A) and critical current $I_c$ vs applied magnetic field at $\alpha =0^\circ$ of a 80-nm thick YBCO film (sample SQ200), irradiated with a slightly defocused He$^+$ ion beam of 50~nm FWHM to form a square pattern of defect cylinders with a lattice constant of 200~nm. Data were taken after zero-field cooling and then sweeping the field through a full cycle, revealing no hysteresis. The matching field determined from the geometric parameters is $B_1 = 52$~mT and leads to a minimum of the resistance and a peak in the critical current.}
\label{fig:R_Ic_200}
\end{figure}

In electric transport measurements, the commensurability effects evoked by regular pinning lattices are demonstrated in Fig.~\ref{fig:R_Ic_200} as peaks in the critical current $I_c$ and corresponding minima of the resistance vs applied field $B_a$ (at $\alpha = 0^\circ$) that appear exactly at the matching fields

\begin{equation}
\label{eq:matching}
B_n = n \frac{{{\Phi _0}}}{{{d^2}}},
\end{equation}
where $n$ is a rational number. We use $n = 0$ to denote the absence of vortices and $n<0$ for the reversed vortex orientation. When the diameters of the CDs are larger than the Ginzburg-Landau coherence length, an integer number $n > 1$ of fluxons can be accommodated per CD. \cite{BUZD93} Note the tiny humps of $I_c$ around $\pm 26\,$mT that indicate a fractional matching pattern with $n = \pm 1/2$.

The commensurability effects result from two different vortex pinning mechanisms in our samples. On the one hand, $n$ flux quanta can be trapped in the normal-conducting core of a CD, which we will call fluxons to discriminate them from the regular Abrikosov vortices in a plain superconductor. These fluxons remain pinned at the CDs even if a moderate current is applied to the sample. However, by changing the applied magnetic field, the Lorentz force due to increased shielding current exceeds the pinning potential and the fluxons can hop between neighboring CDs. \cite{SORE17}

On the other hand, vortices at interstitial positions between the CDs are pinned mainly by twin boundaries and growth defects in the YBCO films, most of them oriented parallel to the $c$ axis. \cite{DAM99} Their pinning potentials are usually weaker than those of the fluxons trapped in the CDs. At a certain applied magnetic field $B_a$ the critical current shows a peak when the magnetic flux through the sample is penetrating the sample via single fluxons trapped in each CD, which happens exactly at the matching field $B_1$ of Eq.~\ref{eq:matching}. In this situation, the number of weakly-pinned interstitial vortices is minimized. An equivalent consideration leads to the explanation of the resistance minima observed at the same $B_n$. Typically, our samples patterned by masked or focused He$^+$ ion irradiation show clear matching effects in a temperature range from $\sim 0.7\,T_c$ up to $\sim 0.9\,T_c$.\cite{ZECH17a,AICH19} For our further considerations it is important that the matching fields can be equally well determined from either $I_c$ peaks or resistance minima, the latter allowing for much faster measurements.

An investigation of the angular dependence of the magnetoresistance can shed light on the nature and relative strength of the pinning of fluxons at the CDs and the pinning of interstitial vortices, respectively.

For dominant pinning at CDs the magnetic flux should be preferentially trapped within the CDs irrespectively of the angle $\alpha$ by which the applied magnetic field $B_a$ is tilted off the axes of the CDs. Then, the commensurability peaks in $I_c(B_a)$ and dips in $R(B_a)$ should appear if the component of $B_a$ that is parallel to the axes of the CDs,
\begin{equation}
 {B_{\parallel} = B_a \cos{\alpha}.}
\label{eq:scale1}
\end{equation}
fulfills the matching condition of Eq.~\ref{eq:matching}.

\begin{figure}[t]
\centering
\includegraphics*[width=\columnwidth]{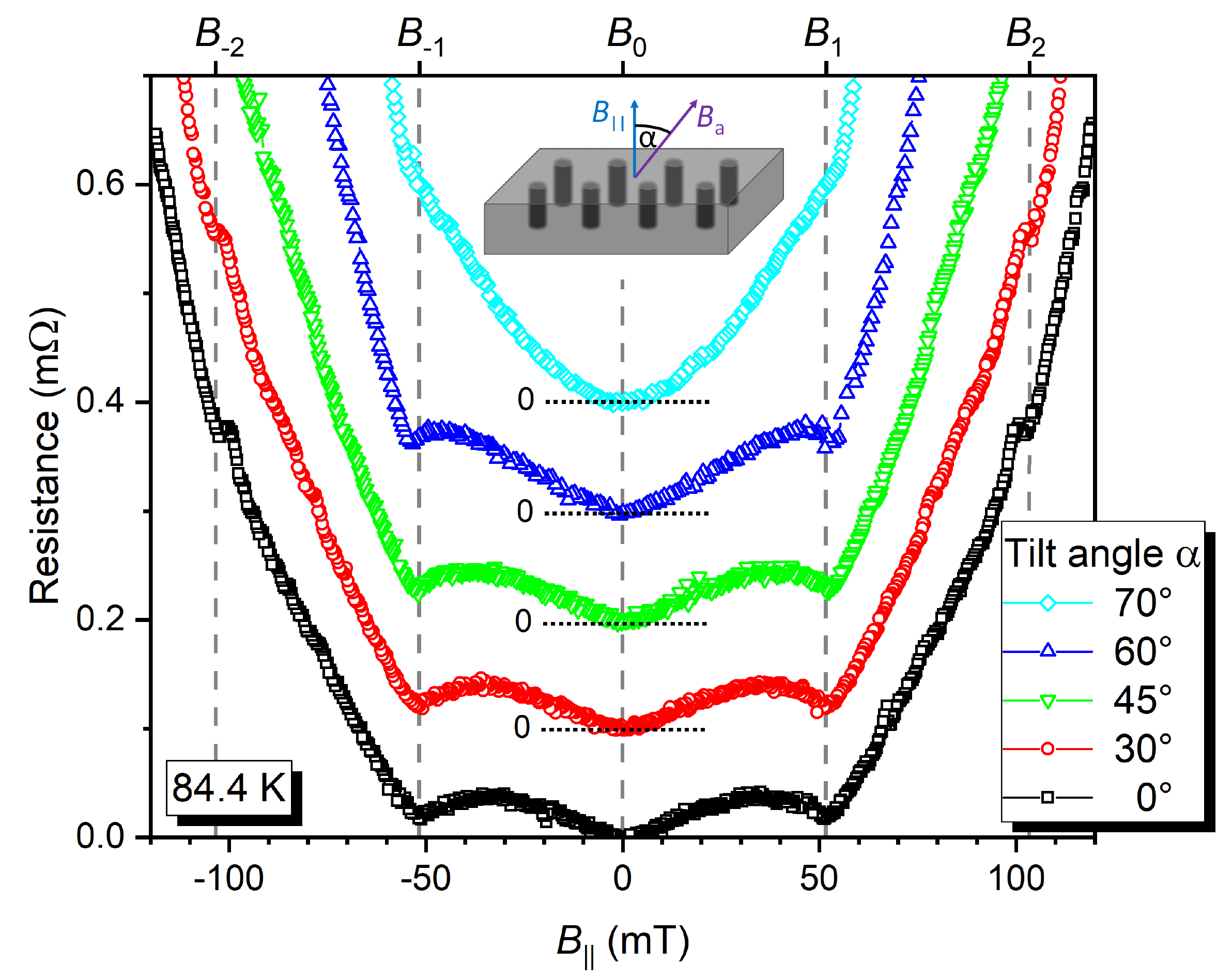}
\caption{Resistance ($I = 400\, \mu$A) vs applied field component along the normal of the film surface $B_\parallel$ of sample SQ200 for different values of $\alpha$.  Since no hysteresis is observed, only the down sweep branches of the cycle  after zero-field cooling are displayed. For $\alpha > 0^\circ$ the curves are shifted by multiples of $0.1\, \text{m} \Omega$ to enhance visibility. The inset shows a sketch of the experimental situation.}
\label{fig:alpha_200}
\end{figure}

Figure~\ref{fig:alpha_200} shows the magnetoresistance of sample SQ200 for various tilt angles $\alpha$ at a temperature near the onset of dissipation. When the magnetic field $B_a$ is oriented orthogonal to the sample surface and parallel to the axes of the CDs ($\alpha = 0^\circ$) a distinct minimum at $B_1 = 52$\,mT and a marginal one at $B_2= 104$\,mT confirms the commensurability effects. With increasing tilt angle $\alpha$ the magnetoresistance curves exhibit very similar matching resistance minima and change their shapes only slightly if data are plotted with the abscissa scaled to  $B_\parallel$. Even at $\alpha = 70^\circ$ the commensurability effect can be detected.

\begin{figure}[t]
\centering
\includegraphics*[width=\columnwidth]{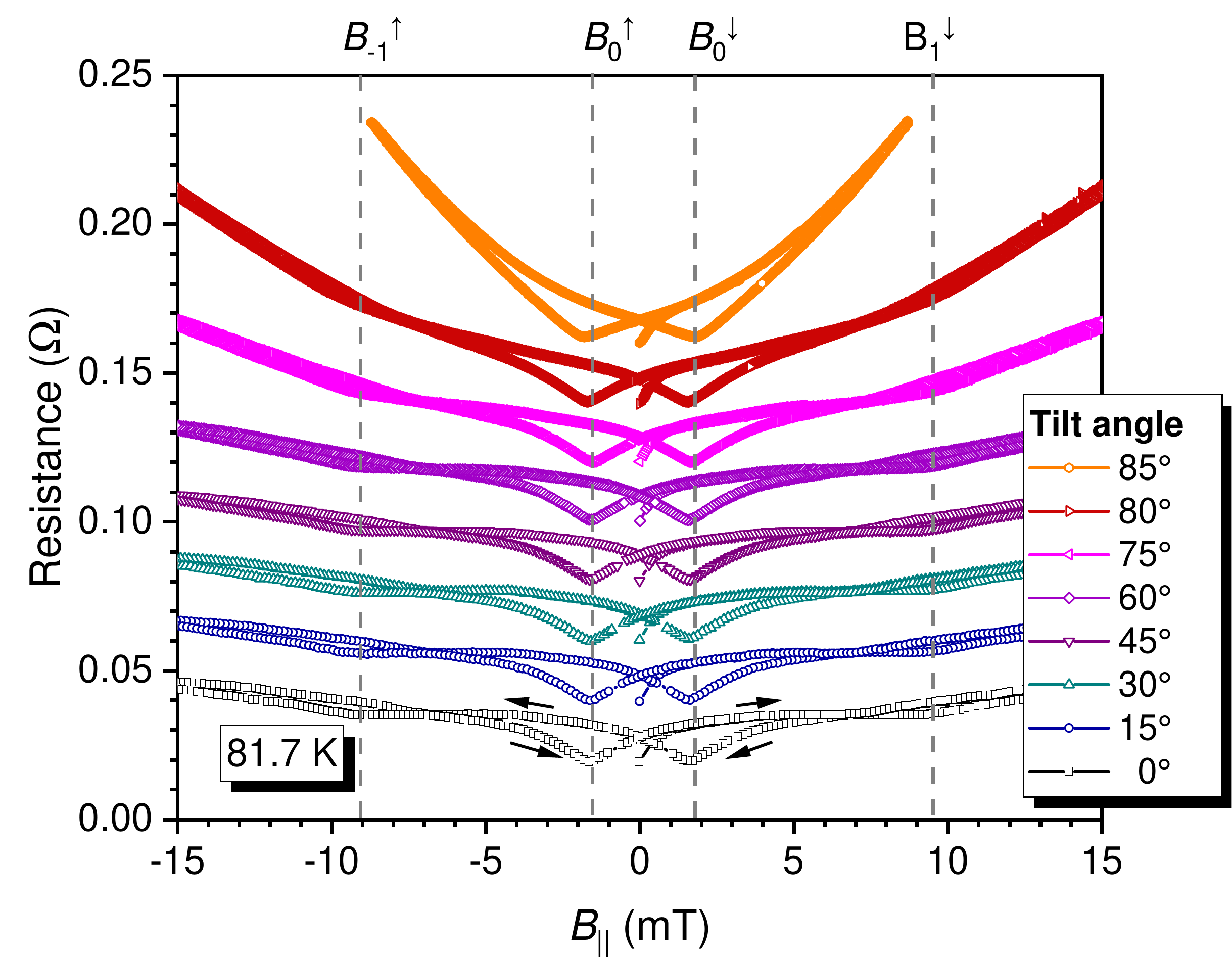}
\caption{Resistance ($I = 200\,\mu$A) vs applied field component along the normal of the film surface $B_\parallel$ of sample SQ500 for different values of $\alpha$. Data were taken after zero-field cooling and comprise the virgin curves starting from $B=0$ and the up and down sweeps as representatively indicated by arrows in the bottom curve. Data for $\alpha > 0^\circ$ are shifted by multiples of $0.02\, \Omega$.}
\label{fig:alpha_500}
\end{figure}

In sample SQ500 the situation is more complicated due to a hysteresis observed in the magnetic field sweeps. It originates from an unconventional terraced critical state \cite{COOL95} with domains in the sample, \cite{REIC97} inside which the pinning centers are occupied by the same number $n$ of fluxons and neighboring domains by $n \pm 1$. Such a hysteretic behavior has been investigated previously \cite{ZECH17a} and is beyond the scope of this work. Still, the considerations leading to Eq.~\ref{eq:scale1} should hold. Indeed, Fig.~\ref{fig:alpha_500} demonstrates that all the features observed in the $\alpha = 0^\circ$ orientation of $B_a$ appear at the same positions when the magnetic field is tilted and scaling to $B_\parallel$ is used. This not only applies to the first matching fields in upsweep ($B_{-1}^\uparrow$) and downsweep ($B_1^\downarrow$) conditions, but also to the hysteretic displacement of the minima with zero fluxon occupation of the relevant CDs ($B_0^\uparrow$ and $B_0^\downarrow$). Despite of the more complex fluxon arrangements in this sample, all commensurability effects are governed by $B_\parallel$, which confirms that only the component of the magnetic field is relevant that is parallel to the axes of the CDs.

\begin{figure}[t]
\centering
\includegraphics*[width=\columnwidth]{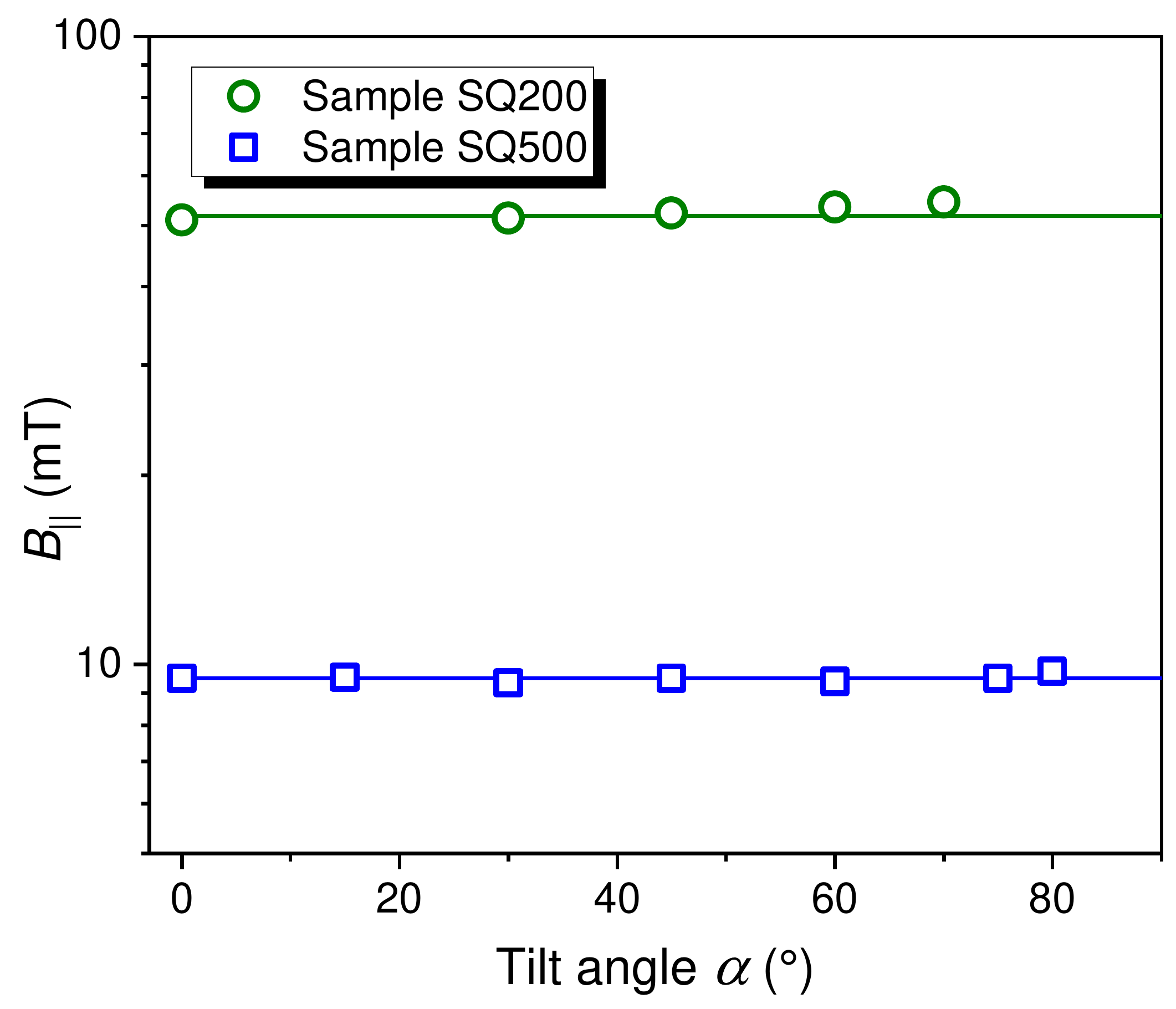}
\caption{Angular dependencies of the magnetic field components $B_{\parallel}$ at which the resistance dips for single fluxon matching $B_1$ (sample SQ200) and $B_1^\downarrow$ (sample SQ500) are observed. The horizontal lines indicate that $B_{\parallel}$  determines the matching effect, irrespective of the tilt angle $\alpha$.}
\label{fig:B1}
\end{figure}

In Fig.~\ref{fig:B1} the magnetic field components $B_{\parallel}$ at which the resistance dips for single fluxon matching are observed in sample SQ200 ($B_1$) and SQ500 ($B_1^\downarrow$) are shown as a function of the tilt angle $\alpha$. In remarkable agreement with Eq.~(\ref{eq:scale1}) the experimental values are independent of $\alpha$ as indicated by the horizontal lines. This confirms that at all angles shown in the graph the magnetic flux is penetrated along the CDs. In addition, the adherence to Eq.~(\ref{eq:scale1}) up to large tilt angles indicates that pinning at the CDs is much stronger than the intrinsic pinning of interstitial vortices in the intermediate regions between the CDs.

Some deviations from the behavior presented in Fig.~\ref{fig:B1} have been reported in denser pinning lattices. Due to lateral straggling of the collision cascades, a significant number of irradiation defects are created in the spaces between the CDs. This is indicated by $\Delta T_c = 43$\,K after MIBS irradiation. In this case the scaling according to Eq.~\ref{eq:scale1} gradually breaks down for $\alpha > 45^\circ$. \cite{ZECH18} In thin YBCO films patterned via 110~keV O$^+$ ion irradiation ($\Delta T_c \simeq 40$\,K) a strong modification of the vortex-glass transition and a weakening of the vortex correlations along the $c$ axis has been observed. \cite{TRAS13}

Finally, in unirradiated YBCO, due to its anisotropy, the cylindrical vortices change to an elliptical cross section in oblique magnetic fields $\alpha \neq 0^\circ$ and decompose into a tilted stack of pancake vortices at tilt angles $\alpha \gtrsim 54^\circ$. \cite{BLAT94R} This is reflected by a broad maximum in the critical current extending over a range $\alpha \lesssim 60^\circ$. \cite{ROAS90} The feature evolves at temperatures closer to $T_c$ and in moderate magnetic fields. In contrast to the observations in those unirradiated YBCO films, the matching fields $B_1$ in our samples strictly scale with Eq.~(\ref{eq:scale1}) up to $\alpha = 72^\circ$ ($\alpha = 80^\circ$) for sample SQ200 (SQ500). Naturally, no pinning of fluxons by the CDs  is expected when the CDs and $B_a$ are oriented orthogonally, i.e., near $\alpha = 90^\circ$.

\section{Conclusions}

Vortex pinning landscapes in YBCO thin films can be conveniently fabricated by employing He$^+$ ion irradiation, either by a focused beam in a HIM or by shadow-masking of a wide-field ion beam. As demonstrated by simulations of the defect distributions created in YBCO by the ion impact, the methods are complementary. HIM irradiation is a sequential method and allows for maskless operation and higher resolution of at least 10\,nm,\cite{MULL19} but the penetration depth is limited to about 80~nm due to the maximum ion energy of 30~keV. With MIBS the entire pattern can be prepared at the same time and also in thicker films when using higher ion energies, but the lateral resolution is currently limited by a hole diameter of $\sim 180\,$nm of the available stencil masks.

Both methods, despite of their different length scales, produce well-defined CDs that provide strong pinning of fluxons, which is supported by the observation that at arbitrary angles of an applied magnetic field only the component parallel to the CDs governs the commensurability effects. Both irradiation methods appear suitable for the creation of well-defined tailored pinning landscapes in cuprate superconductors, which are an important prerequisite for proposed concepts of fluxon manipulation leading to fast and low-dissipation devices. \cite{WAMB99,HAST03,MILO07}

\section*{Acknowledgments}
Research was conducted within the framework of the COST Action CA16218 (NANOCOHYBRI) of the European Cooperation in Science and Technology. B.M. acknowledges funding by the German Academic Scholarship Foundation. M.D. acknowledges the European Erasmus Mundus (Target II) program for financial support.

\end{document}